\documentclass[pra,reprint,showpacs,superscriptaddress,floatfix,nofootinbib]{revtex4-1}
\usepackage{epsfig}
\usepackage{amsmath}
\usepackage{amssymb}
\usepackage{amsfonts}
\usepackage{mathptmx}
\usepackage{dcolumn}
\usepackage{eucal}
\usepackage{bm}
\usepackage{color}
\usepackage[colorlinks,linkcolor=blue,citecolor=blue]{hyperref}
\usepackage{epstopdf}
\usepackage{bbold}
\usepackage{url}
\usepackage{slashed}
\usepackage{mathtools}
\usepackage[dvipsnames]{xcolor}
\usepackage{dsfont}
\usepackage[normalem]{ulem}
\usepackage{framed}

\newcommand{\de}{\partial}

\def \ket#1{\mathinner{|{#1}\rangle}}

\newcommand{\ketbra}[2]{{\mathinner{| {#1} \rangle \langle {#2} |}} }
\newcommand{\matrixel}[3]{{\mathinner{\langle{#1}| {#2} | {#3}\rangle}} }

\newcommand{\lapseApp}[1]{N}

\newcommand{\Avec}{\textbf{A}}
\newcommand{\nablavec}{\boldsymbol{\nabla}}
\newcommand{\Bvec}{\textbf{B}}
\newcommand{\sigmavec}{\boldsymbol{\sigma}}
\newcommand{\taux}{t}
\newcommand{\xix}{z}

\begin{document}

\title{On the Measurement of the Unruh Effect  Through Extended  Quantum Thermometers}

\author{M. Cardi}
\affiliation{Dipartimento di Fisica, Universit\`a di Genova, via Dodecaneso 33, I-16146, Genova, Italy}
\affiliation{INFN - Sezione di Genova, via Dodecaneso 33, I-16146, Genova, Italy}
\author{P. Solinas}
\affiliation{Dipartimento di Fisica, Universit\`a di Genova, via Dodecaneso 33, I-16146, Genova, Italy}
\affiliation{INFN - Sezione di Genova, via Dodecaneso 33, I-16146, Genova, Italy}
\author{N. Zangh\`i}
\affiliation{Dipartimento di Fisica, Universit\`a di Genova, via Dodecaneso 33, I-16146, Genova, Italy}
\affiliation{INFN - Sezione di Genova, via Dodecaneso 33, I-16146, Genova, Italy}

\date{June 13, 2024}

\begin{abstract}
The Unruh effect, predicting a thermal reservoir for accelerating systems, calls for a more refined understanding of measurement processes involving quantum systems as thermometers. 
Conventional models fail to account for the inherent spatial extent of the thermometer, neglecting the complexities associated with accelerated extended quantum systems. Our work builds upon the seminal work of Bell, Hughes, and Leinaas \cite{Bell1985}. We propose a refined thermometer model incorporating a spin-1/2 particle where the spin acts as a temperature indicator. This refined model demonstrates the ability to effectively measure the temperature under specific, realistic conditions, providing a unique value that essentially averages the local Unruh temperatures throughout the extended quantum system acting as the thermometer.
\end{abstract}

\maketitle

\section{Introduction}
\label{sec:intro}

The Unruh effect arises from the predictions of relativistic quantum field theory and has a direct connection to the phenomenon of Hawking radiation emitted by black holes.  Discovered independently by Fulling, Davies, and Unruh \cite{Fulling1973, Davies1975, Unruh1976} in the early Seventies, and later further clarified by DeWitt \cite{DeWitt1979}, the effect describes a counterintuitive consequence of constant proper acceleration: the measurement of a finite temperature in a seemingly empty vacuum.  
If the proper acceleration is denoted by $a$, the corresponding Unruh temperature $T_a$  reads 
\begin{equation}
    T_a = \frac{\hbar a}{2 \pi c k_B}\,, 
    \label{eq:Unruh_temp}
\end{equation}
where $\hbar$, $k_B$, and $c$ are Planck's constant, Boltzmann's constant, and the speed of light, respectively. However, the Unruh effect goes beyond this simple formula. While the theory is rooted in well-established physics, directly observing it has proven difficult. Our present work investigates the measurability of the Unruh temperature and aims to determine the status of the Unruh effect as a physical phenomenon revealed by suitable quantum thermometers.

Observing the Unruh effect remains challenging due to the extreme accelerations required (on the order of $a \approx 10^{20}$ m/s$^2$) to produce an Unruh temperature of even one Kelvin.
As such accelerations are currently beyond practical reach, alternative experimental setups have been proposed.
Among these,   one of the most promising involves utilizing storage rings where particles can be accelerated close to the speed of light \cite{Bell1983, Bell1987, Leinaas2002}.
However, it has been shown that also in this scenario, unequivocally identifying the effect presents challenges \cite{Bell1983,Bell1987}.

Recently, advancements in quantum technology, enabling manipulation and measurement at unprecedented precision, alongside notable progress in particle measurements in storage rings \cite{abi2021} have sparked renewed interest in the Unruh effect.
This renewed attention has prompted the development of new experimental proposals aimed at measuring the effect. These proposals encompass various approaches, including utilizing Berry phase to enhance the detector sensibility \cite{Martin-Martinez2011}, utilizing particular set-ups to increase the light-matter coupling  \cite{Soda2022}, and employing high precision and optimized cavities acting to enhance the excitation rates \cite{Stargen2022}.
At the same time, set-ups for the detection of analogous effects for condensed-matter systems have been proposed  \cite{Gooding2020}.

This resurgence of interest in the Unruh effect prompts us to reconsider the very nature of measurements involved in detecting it.
The most used models describe the quantum systems moving along classical localized paths. While this approach offers simplicity, it deviates from realistic implementation and theoretical considerations.
Conventional analyses often overlook intricate quantum effects arising when measurement devices themselves possess significant physical size. Bucholtz and Verch \cite{Buchholz_2015} emphasized this perspective and acknowledged the pioneering work of Bell, Hughes, and Leinaas (BHL) in 1985  \cite{Bell1985} who explored extended thermometers for the Unruh effect. Subsequent studies have further addressed this issue \cite{Korsbakken_2004}, \cite{Lima2019}, \cite{Foo2020}, and \cite{Barbado2020}. 

Building upon the BHL framework, we address a key challenge: devising suitable quantum measurements to characterize the thermal behavior of accelerated quantum systems. This challenge arises because the concept of a single Unruh temperature becomes ambiguous for extended systems. Our work explores how measurements can be designed to account for this ambiguity, paving the way for a more comprehensive understanding of thermal behavior in these systems.
  
We begin with a general discussion on using an extended quantum system as a thermometer for the Unruh effect, independent of the system's specific nature. Subsequently, we focus on a specific case: a single electron confined within a box by external fields. This electron is treated as a low-energy excitation of a zero-temperature fermionic Dirac field in the Minkowski vacuum. Similar to BHL, we consider a scenario where these external fields not only confine the electron but also uniformly accelerate it.
A crucial aspect of this approach involves a suitable boosting of the quantum state at a constant rate determined by the acceleration $a$. This boosting procedure, which will be explained in detail later, accounts for the effects of uniform acceleration on the electron's quantum state. Additionally, the fermionic field interacts with a surrounding bosonic field, such as the electromagnetic field, in its vacuum state. 

We extend the BHL analysis by considering a quantum state for the electron delocalized across a relatively large confinement region. The fully relativistic approach (entirely quantum, without classical trajectories) offers several advantages.  These advantages include the ability to systematically account for approximations, thereby clarifying the underlying physics and providing a complete picture of all relevant relativistic effects. Within this robust framework, we delve into a regime where a non-relativistic treatment suffices. Our focus lies on determining the stationary density matrix of the electron's spin, which serves as our temperature indicator. Interestingly, our analysis reveals that, in general, this stationary state deviates from a thermal state. (For somewhat related investigations, see \cite{Foo2020, Barbado2020}.)

However, our analysis goes beyond this initial finding. It suggests the possibility of the electron's spin experiencing Unruh thermalization even when its quantum state is delocalized. We explore the conditions for this to occur, focusing on physically relevant cases: extended confining potentials and sharply localized wells. Notably, in these scenarios, the measured temperature retains a signature of the system's size. This translates to an average inverse Unruh temperature, reflecting the non-local nature of the extended quantum system.

The paper is organized as follows.
Section~\ref{sec:model} establishes the foundation for our investigation. Here, we revisit the core principles of a uniformly accelerated system, delving into its geometry and quantum dynamics. We then introduce the model's fully relativistic microscopic Hamiltonian.  A key challenge addressed in this section is the ambiguity associated with defining the Unruh temperature in the context of extended quantum systems. Finally, we establish the regime where an approximate, non-relativistic description becomes sufficient.
Section~\ref{sec:accelerated_thermo} investigates the electron's spin as a potential temperature indicator in a situation where its spin degrees of freedom are coupled to its spatial degrees of freedom. The section determines the stationary density matrix of the spin and explores two relevant confining potential scenarios to understand the system's behavior under different confinement conditions. Section~\ref{sec:conclusions} summarizes the key findings of this work and discusses their implications.

\section{The Unruh Effect in Extended Thermometers}
\label{sec:model}

In this section, we revisit and extend the methods and findings outlined in BHL's work \cite{Bell1985}, focusing on employing accelerated extended thermometers to measure the Unruh effect.

\subsection{Accelerated  Frames as Sequences of Inertial Frames}
\label{sec:frame}
In special relativity, the rest frame of an extended object undergoing constant proper acceleration while maintaining Born rigidity (meaning it experiences no internal stresses or deformations) is typically described using Kottler-Møller (KM) coordinates or Rindler coordinates. As we will recall, these coordinates can conveniently be viewed as a sequence of instantaneous inertial frames. This aspect simplifies the description of accelerated extended quantum systems.

Starting with inertial coordinates \((cT, X, Y, Z)\) in Minkowski spacetime and assuming motion along the \(Z\) direction, the KM coordinates \((ct, x, y, z)\) are given by \cite{Moller1972,Kottler1914}
\begin{align}
\begin{aligned}
    cT &= \left(z + \frac{c^2}{a}\right)\sinh{\left(\frac{a t}{c}\right)} \\
    Z &= \left(z + \frac{c^2}{a}\right)\cosh{\left(\frac{a t}{c}\right)} - \frac{c^2}{a}
    \\
    X &= x, \quad Y = y
\end{aligned}
\label{eq:cotra}
\end{align}

\begin{figure}
    \begin{center}
    \includegraphics[scale=.6]{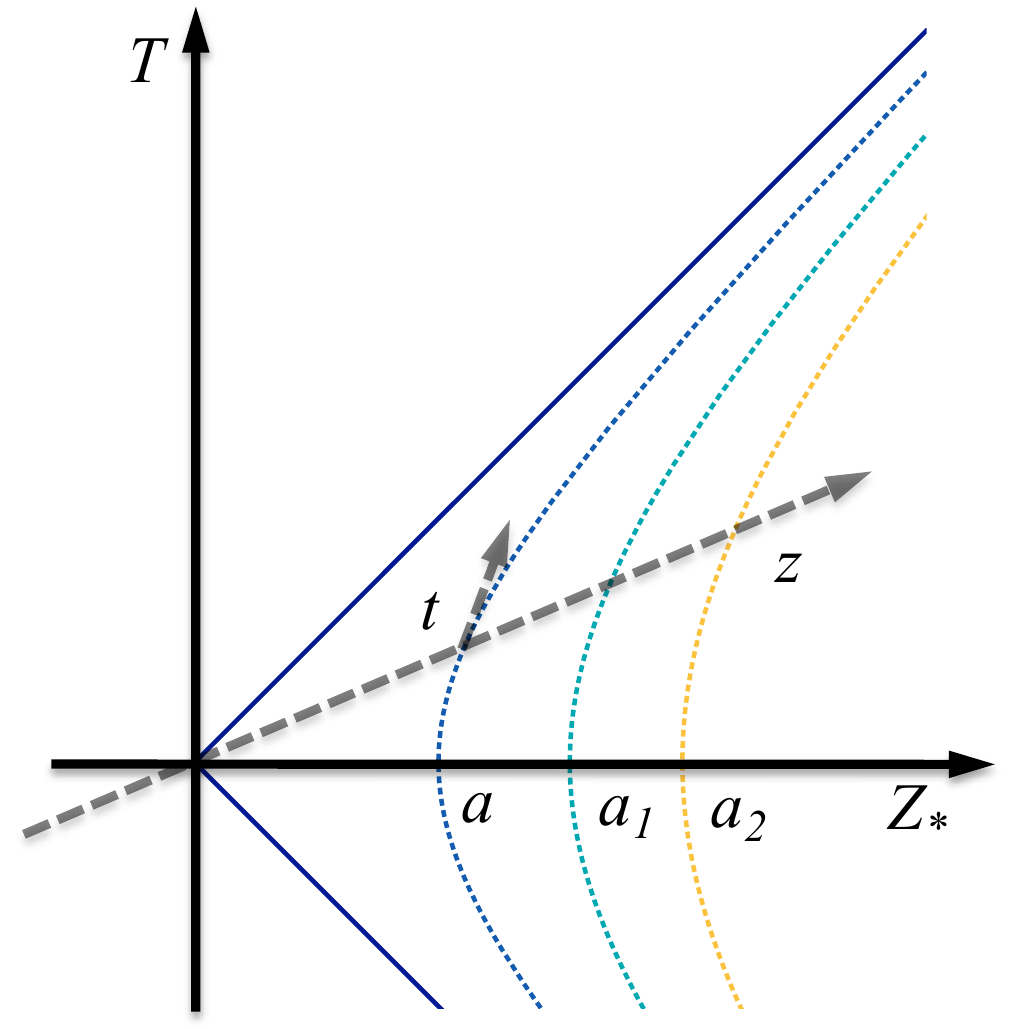}
       \end{center}
    \caption{Coordinates in the  accelerated frame: $Z_* = Z+c^2/a$ and $T$ are the Minkowski coordinates while $\taux $ and $\xix$ are Kottler-M\o{}ller coordinates in the Rindler frame. The hyperbolas in the picture represent the trajectories of bodies at rest in the Rindler frame, with respect to the global time $\taux $, and have different accelerations, say $a$ (corresponding to $\xix=0$),  $a_L$ and $a_R$.
} 
    \label{fig:fig1}
\end{figure}

The pair \((z+  {c^2}/{a}, t)\) is commonly referred to as Rindler coordinates (see Fig. \ref{fig:fig1}). The Kottler-Møller (KM) coordinates, however, do not cover the entirety of spacetime but uniquely parametrize the so-called right Rindler wedge \( Z + {c^2}/{a} > |T| \) and the left Rindler wedge \( Z +  {c^2}/{a} < |T| \). Without loss of generality, we shall henceforth consider motion only in the right Rindler wedge.

In KM coordinates, the Minkowski metric takes the form
\begin{align}
ds^2 = -c^2N ^2 d t ^2 + dx^2 + dy^2 + dz^2\,.
\label{eq:metro}
\end{align} 
where \begin{align} 
N(\xix) = 1+\dfrac{a\xix}{c^2} ,
\label{km} 
\end{align} is the so-called ``lapse function'' which determines the local time dilation at $(x, y, z)$ during the interval $dt$, with a slower rate of time variation for $z > 0$ in the comoving frame. In particular, a motion with frequency $\omega_0$ exhibits a frequency $\omega = N(\xix)^{-1}\omega_0$ in the comoving frame, indicating red-shift for $z > 0$.

Note that the parameter \(a\) in the coordinate transformations in Eq. \eqref{eq:cotra} represents the proper acceleration of motion in the \(Z\)-direction, passing through the origin \(X=Y=Z=0\) at \(T=t=0\). The global time \(t\) in the KM coordinates corresponds indeed to the proper time of this motion, described by the hyperbola
\begin{align}
\label{rindlercoord}
\begin{aligned}
    cT(t) &= \frac{c^2}{a}\sinh\left(\frac{at}{c}\right) \\
    Z(t) &= \frac{c^2}{a}\left[\cosh\left(\frac{at}{c}\right) - 1\right] \\
    X &= Y = 0\,.
\end{aligned}
\end{align}

One obtains a foliation of the right Rindler wedge into hyperbolas by varying \(z\) in Eq. \eqref{eq:cotra} (similarly for the left Rindler wedge). The proper acceleration of the hyperbola passing through \(z\) at \(t=0\) is given by:
\begin{equation}
\label{eq:propacc}
    a(z) = \frac{a}{N(z)}
\end{equation}
and the proper time along this hyperbola is:
\begin{equation}
    \label{eq:proptime}
    \tau(z) = N(z) t.
\end{equation} 
This relationship ensures that \(a(z)\tau(z) = at\) holds for all hyperbolas, guaranteeing that different parts of an extended body momentarily at rest in the KM coordinates have an acceleration profile so that the distance between the different points of the body remains constant in their proper frame. Thus, as recognized long ago, the KM coordinates provide a frame well-suited for describing an accelerated extended ``rigid'' body.

Of particular relevance for subsequent discussions is the following observation: Without loss of generality, let's confine the discussion to the \((cT, Z)\)-plane and examine the motion of a point-like body described by Eq. \eqref{rindlercoord}. At any given moment, the accelerated body is at rest in a momentary inertial frame. The sequence of these momentary inertial frames corresponds to a successive application in the plane of Lorentz transformations $\Lambda(t)$, with rapidity $at/c$,  from the momentary frame to the original inertial frame,
\begin{equation}
\Lambda(t) =
\begin{pmatrix}
    \cosh(at/c) & \sinh(at/c) \\
    \sinh(at/c) & \cosh(at/c)
\end{pmatrix}\,.
\label{eq:LT}
\end{equation}
 We have:
\begin{equation}
\begin{pmatrix}cT\\ Z\end{pmatrix} = \begin{pmatrix}cT(t)\\ Z(t)\end{pmatrix} + \Lambda (t)  \begin{pmatrix}0\\ z\end{pmatrix}\,,
\label{eq:LTZ}
\end{equation}
leading to the transformation laws \eqref{eq:cotra}. Hence, the KM coordinates can be interpreted as describing events in a sequence of inertial frames \(\mathcal{F}(t)\), parametrized by \(t\) over \(-\infty < t < \infty\), such that the relation between the coordinates of the same event in two infinitesimally close inertial frames \(\mathcal{F}(t)\) and \(\mathcal{F}(t+dt)\) are:
\begin{align}
 \begin{aligned}
    cT_{t+dt} &= cT_t -(a\, dt/c) Z_t - cdt\\
     Z_{t+dt}  &= Z_t - (a\, dt) T_t\,.
     \end{aligned}
    \label{eq:infintesimal}
 \end{align}
In other words, two infinitesimally close members of the inertial frames set \(\mathcal{F}(t)\) and \(\mathcal{F}(t+dt )\) can be transformed into each other through a combination of an infinitesimal time translation and an infinitesimal boost in the \(Z_t\)-direction with infinitesimal rapidity $(a/c)  dt$. This observation will play an important role in the subsequent description of an extended quantum system.

\subsection{Quantum Dynamics in  an Accelerated Frame}
\label{sec:KMS}

We can describe an extended relativistic quantum system using appropriate quantum fields within the framework of quantum field theory, without the need, for the moment, to delve into specifics. Despite evidence to the contrary, it is important to appreciate that describing its dynamics in an accelerated frame doesn't necessitate any formulation of quantum field theory in KM coordinates; traditional tools of relativistic quantum field theory within inertial frames are sufficient. Indeed, in light of the observation around Eq. \eqref{eq:infintesimal}, to describe the dynamics of the system, we only need to consider the operators on the Hilbert space representing the generators of time translations and boosts, as elaborated below.

Let $\Phi(t)$ denote the Heisenberg state of the system in the inertial frame \(\mathcal{F}(t)\). 
Then, in agreement with Eq. \eqref{eq:infintesimal},  $\Phi(t+dt)$ is obtained from $\Phi(t)$  
with a generator that combines the Hamiltonian operator ${H}$ (the generator of time translations in the inertial frame) and the boost operator $\text{K}\equiv \text{K}_z$ along the direction $z$: 
\begin{align}
    \Phi(t+dt) = \bigg[1 - \frac{i}{\hbar}\left( \text{H}  + \frac{a}{c}  \text{K}\right) dt\bigg] \Phi(t)\,.
\end{align}
(Because the acceleration is uniform, the operators $\text{H}$, $\text{K}$ do not explicitly depend on time.)
Then, the operator defined as \begin{equation}\label{eq:kthb}
     \text{H}^a =  \text{H}  + \frac{a}{c}  \text{K}\,.
\end{equation}
can be regarded as the generator of dynamics in the instantaneous accelerated reference frame labeled by $a$. We shall refer to it as the ``boosted" Hamiltonian.

Let us now specify the system further and consider it as comprising a subsystem acting as a `thermometer' interacting with a quantum field, which acts as a `reservoir' in the sense of statistical mechanics. 
For the thermometer, we can consider a quantum fermionic field (as we shall do later). As for the reservoir, we can consider either the quantum electromagnetic field or a bosonic scalar field; in both cases, no specific choice is necessary at this point. 

The  boosted Hamiltonian for this compound system is given by 
\begin{equation} \label{eq:kwuv}
   \text{H}^a = \text{H}^a_\text{T} + \text{H}^a_\text{R} + V 
\end{equation}
where $\text{H}^a_\text{T}$ refers to the thermometer, $\text{H}^a_\text{R}$ to the reservoir it interacts with, and $V$ represents the interaction energy operator. Note that $\text{H}^a_\text{T}$ and $\text{H}^a_\text{R}$ share the same structure as the boosted Hamiltonian \eqref{eq:kthb}, with each one incorporating its inertial Hamiltonian and boost generator, respectively (for a concrete example, see Eq. \eqref{WHB} below).

The interaction between the thermometer and the reservoir is represented by the operator $V$ of the general form   
\begin{align}
\label{eq:vtsum}
    V  = \sum_m a_m  b_m  .
\end{align}
where $a_m $ and $b_m $ are suitable operators associated with the thermometer and the reservoir,  respectively. Moreover, it is assumed that ``initially"  (assuming adiabatic switching on of the interaction), the state of the system is of the form $ \ket{i}\otimes\ket{0}$, where $\ket{i}$ denotes the initial state of the thermometer, and $\ket{0}$ is the Minkowski vacuum state of the fields describing the reservoir.

\subsection{KMS Condition and Thermalization}

Let's begin by reviewing the situation, not in an accelerated frame, but in an inertial frame, and consider the reservoir alone,  say an electromagnetic field in any such frame. Let \( \text{H}_\text{R}\) be its Hamiltonian (in agreement with our preceding notations). In this frame, its thermal equilibrium state at an inverse temperature \( \beta \) is represented by the canonical density matrix \( \rho_\beta \propto \exp\left(-\beta \text{H}_\text{R} \right) \), with the inverse of the partition function \( Z= \text{Tr } \left[\exp\left(-\beta \text{H}_\text{R} \right)\right] \) as proportionality constant.

The expectation value of a product of local Heisenberg operators $A(Z,T)$ and $B(Z',T')$ at times $T$ and $T'$ and spatial coordinates $Z$ and $Z'$ in the inertial rest frame (with coordinates $X$ and $Y$ omitted for simplicity) is
\[\langle A(Z,T) B(Z',T')\rangle = \text{Tr } \left[ A(Z,T) B(Z',T') \rho_\beta\right]\,. \]
By using the explicit formulas for $\rho_\beta$ and the Heisenberg operators,  it is straightforward to find
\begin{align}
\label{eq:kms}
\langle A(Z,T - i\hbar \beta)B(Z', 0) \rangle= \langle B(Z', 0)A(Z,T)\rangle.
\end{align}
Conversely, a well-established theorem states that if Eq. \eqref{eq:kms} holds for all pairs of reservoir operators, then the reservoir's state must indeed be the canonical density matrix $\rho_\beta$ \cite{Kubo1957,Martin-Schwinger1959}. This implies that Eq. \eqref{eq:kms}, known as the KMS condition, offers an alternate representation of a thermal state at inverse temperature $\beta$.

The basic tenet of the theory of the Unruh effect is the observation that the vacuum state of the quantum fields, serving as the reservoir, appears thermal in the non-inertial accelerated frame described above \cite{Fulling1973, Davies1975, Unruh1976}. Particularly insightful is the formulation and derivation of this result presented by BHL, expressed by the following theorem: Let $\langle \ldots \rangle_0$ denote the expectation value with respect to the Minkowski vacuum state $\ket{0}$ of the quantum fields describing the reservoir. Additionally, consider $A(z,t)$ and $B(z',t')$  as local Heisenberg operators in relation to the dynamics governed by the boosted Hamiltonian $\text{H}^a_\text{R}$. Then,
\begin{align}
\label{acckms}
    \langle A(z, t - i\hbar \beta_a)B(z',0) \rangle_0= \langle B(z',0)A(z,t)\rangle_0
\end{align}
for inverse temperature\begin{equation}
     \beta_a \equiv \frac{1}{k_BT_a} = \frac{2\pi c} {\hbar a} \,.
    \label{eq:Unruh_temp2}
\end{equation}

Various similar results exist in the literature, differing in rigor and generality. For example, for an early rigorous treatment, refer to Ref. \cite{SEWELL1982}, while for a relatively more recent approach and further references, see Ref. \cite{Leinaas2002}. Nonetheless, what distinguishes this result is its simplicity, relying solely on the fundamentals of quantum field theory in inertial frames.

Our interest in Eq. \eqref{acckms} lies in its demonstration of how the Minkowski vacuum generates statistical distributions of Heisenberg quantum observables, evolving over time $t$ in the accelerated frame according to the boosted Hamiltonian $\text{H}_\text{R}^a$. This corresponds to the behaviour expected from the thermal density matrix proportional to $ \exp\left( - \beta_a \text{H}_\text{R}^a\right)$. In essence, within a uniformly accelerated reference frame, the reservoir alone behaves as though it were in a thermal state. (Note that we focus on the reservoir restricted to the right Rindler wedge; similar behavior occurs in the left wedge.)

If the reservoir thermalizes, it's reasonable to expect the thermometer immersed in it to undergo the same process, reaching the same temperature as the reservoir. Specifically, when considering the interaction between the thermometer and the reservoir---the compound system governed by Hamiltonian \eqref{eq:kwuv}---one expects the reduced density matrix of the thermometer to thermalize and become proportional to $ \exp\left( - \beta_a \text{H}_\text{T}^a\right)$.

This result is quite standard in statistical mechanics, typically guaranteed by assuming that the interaction energy described by Eq. \eqref{eq:vtsum} is sufficiently weak and that, in a suitable sense, the reservoir is ``large" compared to the thermometer. Moreover, dealing with quantum fields requires suitable cut-offs to be assumed, particularly to account for discrete energy levels and avoid ultraviolet divergences. Under these assumptions, it can be shown that the transition probabilities $W_{mn}$ from the initial eigenstate $\ket{m}$ to the final eigenstate state $\ket{n}$ of the thermometer's boosted Hamiltonian $\text{H}_\text{T}^a$ satisfy the detailed balance condition
\begin{align}
\label{eq:detailed_balance}
W_{mn} = e^{\beta_a (\mathcal{E}_n-\mathcal{E}_m)}W_{nm},
\end{align}
where $\mathcal{E}_n$, $\mathcal{E}_m$ are the corresponding eigenvalues of  $\text{H}_\text{T}^a$ \cite{Bell1985}.  Accordingly, thermal equilibrium is reached when  the occupation probabilities $p_i$ and $p_f$ of any two states satisfy the condition
\begin{align}
    \dfrac{p_n}{p_m} = \dfrac{W_{nm}}{W_{mn}} = \dfrac{e^{-\beta_a \mathcal{E}_n}}{e^{-\beta_a \mathcal{E}_m}},
    \label{eq:thermal_rho}
\end{align}
i.e., when the thermometer shares the same temperature as the reservoir, and its reduced density matrix becomes proportional to $ \exp\left( - \beta_a \text{H}_\text{T}^a\right)$, as expected.

Equivalently, the same conclusion can be reached using the standard formalism to derive the master equation governing the evolution of a subsystem's reduced density matrix within a larger system \cite{Breuer_book,Blum_book}). By starting with the boosted Hamiltonian \eqref{eq:kwuv} and employing the interaction representation while utilizing standard approximations (e.g., Born, Born-Markov, and secular approximations), one readily obtains the master equation governing the evolution of the reduced density matrix of the thermometer:
\begin{equation}
	\frac{d{\rho}_{mm}}{dt}   =  \sum_{n \neq m} \rho_{nn}   W_{mn} - \rho_{mm} \sum_{n \neq m} W_{nm}.
	\label{eq:rate_eq}
\end{equation}
Here, $\rho_{mn}$ denotes the matrix elements of the reduced density matrix of the thermometer with respect to the eigenstates of $\text{H}_\text{T}^a$. Due to detailed balance \eqref{eq:detailed_balance}, the equilibrium solution of Eq. \eqref{eq:rate_eq} indeed corresponds to the previously identified thermal state of the thermometer.

In summary, when immersed in a quantum field initially in its Minkowski vacuum state, a thermometer subjected to uniform acceleration reaches a thermal state described by the density matrix
\begin{equation}
\rho^a_{\beta_a} \propto \exp\left( - \beta_a \text{H}^a\right)
\label{eq:thermal_rho_a}
\end{equation}
in the instantaneous Lorentz frame. Here, $\text{H}^a$ denotes the boosted Hamiltonian of the thermometer alone. Abusing notations,  we have dropped the subscript ``$\text{T}$'' since we shall exclusively focus only on the thermometer for the remainder of the paper. 

\subsection{Temperature Ambiguity for an  Extend Thermometer}
\label{sec:ambiguity}

Before delving into the potential measurement of the thermalization process outlined earlier through implementing a realistic thermometer model, we pause and comment on the type of thermalization we are dealing with. As we will elucidate further, drawing from the insights of Ref. \cite{Bell1985}, we encounter a form of thermalization that defies comparison within both classical theories of extended relativistic systems and the framework of quantum statistical mechanics applied to extended systems in inertial frames.

In the KM coordinates \eqref{eq:cotra}, let's consider the change of variables:
\begin{align}
\begin{aligned}
\bar{z} &= z - z_0 \\
\bar{t} &= \left(1 + \frac{a}{c^2} z_0 \right) t \\
\bar{a} &= \left(1 + \frac{a}{c^2} z_0 \right)^{-1} a
\end{aligned}
\end{align}
where $z_0$ is some given fixed value. This transformation yields:
\begin{align}
\begin{aligned}
cT &= \left(\bar{z} + \frac{c^2}{\bar{a}}\right) \sinh{\left(\frac{\bar{a} \bar{t}}{c}\right)} \\
Z &= \left(\bar{z} + \frac{c^2}{\bar{a}}\right) \cosh{\left(\frac{\bar{a} \bar{t}}{c}\right)} - \frac{c^2}{\bar{a}} + z_0.
\end{aligned}
\label{eq:cotra2}
\end{align}
Consequently, except for a trivial shift in the $Z$-variable, the transformation laws remain unchanged. This implies that for a given extended system in motion, attributing a different ``uniform" acceleration $\bar{a}$ instead of $a$ is equally valid, achieved simply by choosing $z = z_0$ rather than $z = 0$ as the ``center" of the system. Consequently, applying the same reasoning as before, we arrive at a distinct Unruh temperature
\begin{equation}
T_{\bar{a}} = \frac{\bar{a}}{a} T_a
\end{equation}
This ambiguity challenges the very notion of the Unruh temperature of an extended system.

However, if the system is ``small", i.e., the quantum system is strongly localized around some point $z=z_0$ (and thus remains in the accelerated frame for a sufficient amount of time), selecting a ``central point" within the system is straightforward, leading to the conclusion that the system is characterized by a single temperature. Conversely, in the case of an extended system, we could choose any point inside the quantum system as the ``central point". And since different points are associated with different proper accelerations, as expressed by Eq. \eqref{eq:propacc}, it becomes impossible to identify a single Unruh temperature. This implies that the very concept of the Unruh temperature of an extended system is ambiguous.

This ambiguity in temperature arises from an inherent uncertainty in both the unit of time and, consequently, the unit of energy. Notably, standard clocks attached to different points in the ``uniformly" accelerated object lose synchrony. Consequently, we could, in fact, define a local temperature $T(z)\sim a/(1+az/c^2)$ associated, at each point $z$. This local temperature would then vary across the object.

Some might argue that this scenario is not unprecedented: a similar situation arises for a classical object in thermal equilibrium within a static gravitational field. In such cases, the presence of gravitational redshifts introduces variations in the natural unit of time and energy between different heights, resulting in a temperature inhomogeneity across the body. Nevertheless, this analogy is not appropriate for our scenario.

Here, we deal with an extended quantum system, where the quantum states are not confined to specific locations but instead encompass the entirety of the system. The equilibrium state of the system is characterized by a distribution over these extended states. According to the modern understanding of thermal equilibrium in large systems, thermal equilibrium is typically a property of the extended states of the system, and under normal circumstances, they are associated with a unique thermodynamic temperature \cite{Goldstein2010,Srednicki1994}.

However, in this context, this is not the case. So, as stressed by BHL, we are dealing here with an ambiguity rather than an inhomogeneity in temperature. This is a novel phenomenon without a classical or quantum analogue.

\subsection{Modeling the Thermometer through Confined  Electrons}
\label{sec:hamiltonian}

Given the challenges mentioned earlier, it's natural to wonder if we can actually measure the thermal effect.
The key lies in identifying the right measurable quantities, similar to a ``temperature indicator", that can represent a sort of average temperature and maybe provide a result independent of any arbitrarily chosen ``central point''.  Choosing this indicator depends on a more detailed model of the system acting as a thermometer, which we'll explore next.

So far, we've left the concept of the thermometer fairly broad. In theory, it could be any quantum system that follows the laws of Lorentz invariance and has a finite size. However, to create a more practical and workable model, we'll follow the approach laid out by BHL, which treats the thermometer as a single confined electron. (This approach simplifies the analysis but sacrifices a more realistic portrayal of a macroscopic thermometer; we will delve into these complexities in future work).

To keep the confined electron moving with constant acceleration, we must apply special external fields, denoted by $A_\mu= (\varphi, \mathbf{A})$. These fields must remain constant across the series of inertial reference frames $\mathcal{F}(t)$ associated with the accelerated motion. Imagine these fields like a constantly moving box that traps the electron within its frame. (Neither the original work by BHL nor this discussion explores the complex engineering required to create and maintain these fields.)

With these external fields in place, the electron's behaviour can be described using a Dirac quantum field, denoted by $\psi$. This field follows specific rules (canonical anticommutation relations) compared to an empty space devoid of particles (Minkowski vacuum).

According to Eqs. \eqref{eq:kthb} and \eqref{eq:kwuv} the boosted Hamiltonian $\text{H}^a$ of such a thermometer is given by the expression
\begin{equation}
\label{WHB}
    \text{H}^a = \text{H} + \dfrac{a}{c}\text{K}
\end{equation}
where, abusing notation, $\text{H}^a$, $\text{H}$ and $\text{K}$ now exclusively refer to the thermometer.
The familiar expression for the fermionic Hamiltonian $\text{H}$ is
\begin{equation}\label{eq:HamDifield}
\text{H} =  \int d^3 x~\, \mathcal{H} = \int d^3x~ \psi^\dagger H^{(1)}\psi\,,
\end{equation}
where $H^{(1)}$ is the first quantization Dirac Hamiltonian:
\begin{align}\label{eq:hamdipr}
    H^{(1)} = -ic\hbar\boldsymbol\alpha\cdot\textbf{D} +  mc^2\beta + e\varphi.
\end{align}
Here, $\textbf{D} = \boldsymbol\nabla + \frac{e}{i\hbar}\textbf{A}$, $m$ is the mass of the electron, $\varphi$ is the electrostatic potential, $\boldsymbol\alpha= \gamma_0\boldsymbol{\gamma}$ and $\beta=\gamma_0$, where $(\gamma_0, \boldsymbol{\gamma})$ are the Dirac matrices. Furthermore, Eq. \eqref{eq:HamDifield} and subsequent formulas rely on established regularization procedures for products of field operators evaluated at the same spacetime point.

The  boost generator along $z$ in the first quantization is
\begin{align}\label{eq:secboost1}
K^{(1)} &= \dfrac{1}{2c}\left(z H^{(1)} + H^{(1)}z\right)
\end{align}
(see, e.g., \cite{thaller}).
Then, according  to the standard rules (see, e.g., \cite{bjorken}), the generator of the boost on the Hilbert space of the quantum field is
\begin{align}
    \text{K} &= \int d^3x \,\psi^\dagger K^{(1)} \psi 
\label{eq:secboost}
\end{align}
Substituting Eq. \eqref{eq:secboost1} into the above equation and evaluating the expression $H^{(1)}z\psi$, where $H^{(1)}$ is defined by Eq. \eqref{eq:hamdipr}, we readily obtain
\begin{align}
    \text{K} &= \dfrac{1}{c}\int d^3x \,\psi^\dagger \left(z H^{(1)} -\dfrac{ic\hbar}{2}\alpha_z\right)\psi \,.
\end{align}

Finally, upon substituting in Eq. 
\eqref{WHB} the expressions for $ \text{H}$ and $ \text{K}$, we arrive at
\begin{align}\label{eq:explicitW}
    \text{H}^a &= \int d^3 x \bigg[N(z)\mathcal{H} - \dfrac{ia\hbar}{2c}\psi^\dagger\alpha_z\psi \bigg],
\end{align}
where, recalling Eq. \eqref{km},  we have grouped the terms proportional to $\mathcal{H}$. Because of the zero of $N(z) = 1 +az/c^2$, we restrict our analysis to the region where $N(z)>0$, i.e., the so-called right Rindler wedge.  Henceforth, by $H_a$, we denote the restriction of the right-hand side \eqref{eq:explicitW} to this specific region.

The diagonalization of $\text{H}^a$ can be brought back to solving the c-number time-independent Dirac equation in the region $N(z)>0$ (see, e.g., \cite{weinberg}):
\begin{align}
    \left[N(z)\left(-ic\hbar\boldsymbol\alpha\cdot\textbf{D} + mc^2\beta + e \varphi \right) - i\dfrac{a\hbar}{2c}\alpha_z\right] \Psi  &= E  \Psi
    \label{eq:c-Dirac}
\end{align}
where $\Psi$ is a Dirac 4-spinor. Without the external fields  $\varphi$ and $\mathbf{A}$, the lowest non-negative energy state is the fermionic Rindler vacuum.
In this case, the term $N(z) = 1 +az/c^2$ in Eq. \eqref{eq:c-Dirac} might cause excitations from negative to positive energy levels, indicating pair production (in alignment with Dirac's initial interpretation of his equation). However,  these excitations can be disregarded if \begin{equation}
\label{eq:regime}
\hbar a/c \ll mc^2\;\;\text{ and }\;\;  az/c^2 \ll 1\,. 
\end{equation} 

Then, the focus can shift solely to the excitation of the additional electron confined within the potential well around $z=0$. Accordingly, within the regime defined by Eqs. \eqref{eq:regime}, and assuming that the external fields $\varphi$ and $\mathbf{A}$ are not excessively strong, the positive energy solutions of Eq. \eqref{eq:c-Dirac} provide the eigenvalues of $\text{H}^a$ that predominantly govern the populations of the energy levels in Eq. \eqref{eq:thermal_rho} in the right Rindler wedge.

Additionally, given the specified conditions, a suitable nonrelativistic approximation suffices (see Appendix \ref{app:non-relativistic_approximation of the Dirac equation}), which further simplifies the problem of diagonalizing $\text{H}^a_r$ to the diagonalization of the 1-particle 2-spinor nonrelativistic Hamiltonian
\begin{align}
\label{eq:BHLham}
    \dfrac{1}{2m}(i\hbar\nablavec + e \Avec)^2  + e\varphi + ma\xix - \dfrac{e\hbar}{2m}\sigmavec \cdot \Bvec \,.
\end{align}

A natural choice for our temperature indicator, as mentioned before, is the electron's spin along the $z$-axis (directly measured using a suitable macroscopic device or by observing indirect effects like depolarization in a beam of identically prepared electrons). In this approximation, in fact, the spin and orbital motions decouple. Consequently, the thermal distribution across the energy levels implies a spin density matrix \cite{Bell1985}
\begin{align}
\label{eq:BHL_rho}
   \exp\bigg(\dfrac{\hbar e}{2m}\dfrac{\sigma\cdot\Bvec}{k_B T_a}\bigg) = \exp\bigg(\dfrac{\pi ce}{ma}\sigma\cdot\Bvec\bigg),
\end{align}
where $T_a$ is the Unruh temperature associated with the (single) acceleration $a$, i.e., the acceleration of the electron near $z=0$.

We conclude that the ambiguity of temperature doesn't pose a significant challenge in this simplified model and within this approximation. The main hurdle becomes the practical difficulty of confining the electron for an extended period.

\section{Extended Thermometers with Spin-Spatial Degrees of Freedom Coupling}
\label{sec:spin-thermometer}

The results obtained using the lowest order of Eq. \eqref{eq:c-Dirac} omit spin-orbit coupling. Consequently, the BHL model decouples the spin from the spatial degrees of freedom, preventing it from registering the effects of varying accelerations within the extended quantum system. This leads to a Unruh temperature that does not reflect the extended nature of the system, as shown by Eq. \eqref{eq:BHL_rho}. 

Recognizing this limitation, BHL suggested the presence of these effects in higher-order approximations. We address this by incorporating the next-order contributions in $1/c^2$ using a more straightforward approach. We assume that the magnetic field $\Bvec$ is strong enough to make the term $(e\hbar/2m) N(z) \mathbf{B} \cdot \sigmavec$ in Eq. \eqref{eq:c-Dirac} the only relevant term of order  $1/c^2$  (once the 4-Dirac spinor is expressed in terms of two 2-spinors and the standard decompositions of the Dirac matrices into the Pauli matrices are adopted; see Appendix \ref{app:non-relativistic_approximation of the Dirac equation}, in particular Eq. \eqref{eq_app:Pauli_eq}).

To ensure the validity of the regime characterized in Sec. \ref{sec:hamiltonian} and avoid the phenomenon of pair production, conditions \eqref{eq:regime} must hold.
The first condition $\hbar a/c \ll mc^2$ (for an electron) implies $a \ll 10^{29}\,\text{m/s}^2$, which is comfortably satisfied by achievable Unruh accelerations, typically on the order of $10^{20}\,\text{m/s}^2$. For this latter Unruh acceleration, the second condition $az/c^2 \ll 1$ requires $z \ll 10^{-3}\,\text{m}$.
This means that we can have a quantum thermometer delocalized over a ``macroscopic" length scale.

To simplify the treatment, we take $\Bvec$ uniform (in the accelerated frame) and direct along $z$, so that the Hamiltonian to be diagonalized becomes
\begin{equation}
    H^a= -\dfrac{\hbar^2}{2m} \frac{\partial^2}{\partial \xix^2} + U(\xix) + m a\xix + N(\xix) \hbar \omega \sigma_z\,, 
    \label{eq:hambas}
\end{equation}
where \(\omega = |e| B/2m\), \(m\) and \(e\) denote the mass and charge of the electron, respectively, and \(\sigma_z\) is the third component of the Pauli matrices \(\boldsymbol{\sigma}\). Since we selected the spin degrees of freedom as the temperature indicator, this Hamiltonian is the natural choice for modeling the accelerated extended quantum thermometer.
Crucially, the presence of the lapse function $N(z)$ in Eq. (\ref{eq:hambas}) distinguishes our treatment from that of BHL \cite{Bell1985}. This inclusion enables the spin to serve as a temperature indicator for a delocalized thermometer, as discussed below.

\label{sec:accelerated_thermo}
\subsection{Tracing out the spatial degrees of freedom}

In the regime we have been considering, the density matrix  \eqref{eq:thermal_rho_a} should be replaced with 
\begin{equation}
\widetilde{\rho}^a_{\beta_a} \propto \exp\left( - \beta_a {H}^a\right)
\label{eq:thermal_rho_a1}
\end{equation}
with $H^a$ given by Eq. \eqref {eq:hambas}. 
As the spin along the magnetic field direction serves as our temperature indicator, we need to focus on the reduced density matrix in spin space. This quantity is obtained by tracing out the spatial degrees of freedom from the density matrix $\widetilde{\rho}^a_{\beta_a}$. Denoting the latter by $\widehat{\rho}^a$, we obtain
\begin{equation}
  \widehat{\rho}^a \propto \mathrm{Tr}_\text{space} \left[\exp\left( - \beta_a {H}^a\right)\right]  =   \sum_n \langle n |    e^{-\beta_a H^a}| n \rangle
  \label{eq:rhoa}
\end{equation} 
where $| n \rangle$ is any orthonormal basis in  Hilbert space of the purely spatial degrees of freedom.

To compute $\widehat{\rho}^a$, we first need a tractable expression for $\exp\left( - \beta_a {H}^a\right)$. This can be readily achieved by making appropriate approximations based on the physical regime of interest.   Recalling  the definition in Eq. \eqref{km},  we expand  $N(\xix) \hbar \omega \sigma_z$ and decompose the total  Hamiltonian $H^a$   (Eq. \eqref{eq:hambas}) into the following terms: $H_{\text{spin}} = \hbar \omega \sigma_z$ (spin term), $H_{\text{int}} = {a\xix}/{c^2}\hbar \omega\sigma_z$ (interaction term), and $H_{\text{space}}$   (implicitly defined by the remaining terms in Eq. \eqref{eq:hambas}).

Given the separation of scales between the particle's mass-energy and the interaction term ($mc^2 \gg \hbar\omega$), we can employ first-order perturbation theory to treat $H_{\text{int}}$  as a small perturbation to $H_{\text{space}} + H_{\text{spin}}$. This allows us to solve for the eigenstates and eigenvalues approximately. To implement this, let us consider first  the  diagonal  representation  
\begin{align}
   \exp\left( - \beta_a {H}^a\right) = \sum_{n , \sigma} e^{-\beta_a \mathcal{E}_n^{\sigma}}\ketbra{\Psi_n^\sigma\rangle |  \sigma}{\sigma| \langle \Psi_n^\sigma } \,,
     \label{eq:spectral_rho_therm}
\end{align}
where $\sigma=\pm 1$, with $\ket{\sigma}$   denoting  the two eigenstates  of $\sigma_z$,  and  $\ket{\Psi_n^\sigma}\ket{\sigma}$  being  the eigenstates of $H^a$  corresponding to  the  eigenvalues $\mathcal{E}_n^\sigma$. 

The first-order approximation in   $\hbar\omega/mc^2$ of the right-hand side of Eq. \eqref{eq:spectral_rho_therm}  is given by (see
Appendix  \ref{app:perturbative_expansion})
 \begin{align} 
    \sum_{n , \sigma} e^{ -\beta_a {E}_n - \overline{\beta}_n \sigma  \hbar\omega} \ketbra{\Psi_n^{\sigma_{(1)}}\rangle |  \sigma}{\sigma| \langle \Psi_n^{\sigma_{(1)}} } \,,
     \label{eq:spect}
\end{align}
where 
\begin{itemize}
    \item $E_n$ are the eigenvalues of
    $H_{\text{space}}$;
    \item $\overline{\beta}_n = \beta_a(\overline{\xix}_n)$, with $\beta_a(\xix)=\beta_a(1+a\xix/c^2)$, and 
    \begin{equation}
    \overline{\xix}_n = \int \xix |\Phi_n(\xix)|^2 d\xix \, 
    \label{eq:xi_n}
\end{equation}
is the mean position of the particle in the $n$-th eigenstate  $ \Phi_n$ of $H_{\text{space}}$; equivalently, since  $\beta_a =\beta_a(\xix)$ is a linear  function, $\overline{\beta}_n$ can be regarded as the mean value of $\beta$ in the  state $\Phi_n$, i.e., 
\begin{equation}
    \overline{\beta}_n =  \int  \beta_a (\xix) |\Phi_n(\xix)|^2\, d\xix \,;
    \label{eq:meanbeta}
\end{equation}
\item $\Psi_n^{\sigma_{(1)}}$ represents the first-order corrections to the eigenstates of $H^a$ (see Eq. \eqref{app_eq:perturbed_eigs} in Appendix \ref{app:perturbative_expansion}).
\end{itemize}

Finally, substituting the approximate expression for   $\exp\left( - \beta_a {H}^a\right)$   (Eq. \eqref{eq:spect})  into Eq. \eqref{eq:rhoa}  and choosing the basis states $|n\rangle$ in Eq. \eqref{eq:rhoa} to be the first-order corrected eigenstates $\ket{\Psi_n^{\sigma_{(1)}}}$, we obtain the reduced density matrix in spin space up to the first order in ${\hbar\omega}/{(mc^2)}$: 
\begin{equation}
\widehat{\rho}^a \propto \sum_\sigma c_\sigma \ketbra{\sigma}{\sigma}\,,
\label{eq:rether}
\end{equation}
where $c_\sigma$ are coefficients given by
\begin{equation}
c_\sigma = \sum_n e^{ -\beta_a {E}_n -  \overline{\beta}_n \hbar\omega \sigma } \,.
\label{eq:coeff}
\end{equation}

Let's summarize the key quantities in Eqs. \eqref{eq:rether} and \eqref{eq:coeff} for clarity and future reference: $\beta_a$ signifies the inverse Unruh temperature associated with the Rindler frame, $\overline{\beta}_n$ represents the mean inverse Unruh temperature in the spatial states $\Phi_n$ (the eigenstates of $H_\text{space}$) as determined by Eq. \eqref{eq:meanbeta}, and $E_n$ corresponds to the energies of the states $\Phi_n$ (the eigenvalues of $H_\text{space}$).

In its present form, the spin state described by Eq. \eqref{eq:rether} is not thermal. 
However, as we shall elucidate in the following, through appropriate adjustments to the potential $U$ and careful consideration of the order of magnitudes involved, we will demonstrate that the spin state described by Eq. \eqref{eq:rether} may closely approximate a thermal state.

\subsection{Nonlocal Unruh effect in an extended well}
\label{sec:not_sharp_well}

The spatial properties of the systems are determined by the confining potential $U(z)$ which ultimately determines the form of the eigenstates $ \Phi_n$ of $H_{\text{space}}$. 

Assuming \(U(z)\) as an infinitely deep well with a linear size \(L\), with its left side situated in $z=0$, and tilted due to the lapse, the eigenvalues \(E_n\) of the spatial Hamiltonian \(H_\text{space}\) can be determined numerically. It turns out that for an acceleration of the order \(a \sim 2.5 \times 10^{20}\) m/s\(^2\), and \(L \sim 10^{-7}\) m, only the ground state of the spatial degrees of freedom is populated.
That is, in the sum over \(n\) in Eq. \eqref{eq:rether}, only the term \(n=0\), corresponding to the ground state of \(H_\text{space}\), contributes. Consequently, the reduced density matrix \(\widehat{\rho}^a\) becomes
\begin{equation}
\widehat{\rho}^a \propto \sum_\sigma e^{ - \overline{\beta}_0 \hbar\omega \sigma } \ketbra{\sigma}{\sigma}.
\label{eq:rho_non-local}
\end{equation}
This is the thermal state of the spin-thermometer at inverse temperature \(\overline{\beta}_0\) with respect to the spin Hamiltonian \(H_\text{spin}\). In this case, the thermometer registers a unique temperature despite experiencing different accelerations.
Still, the thermometer keeps track of the nonlocal features and extensions of the wells since $\overline{\beta}_0$ represents the mean value of the inverse local Unruh temperatures over the full spatial extension of the thermometer in the ground state.
For a highly localized well, Eq. \eqref{eq:rho_non-local} reproduces the BHL result expressed by Eq. \eqref{eq:BHLham}.

In other words, the spin-based thermometer measures an average inverse Unruh temperature, reflecting the influence of the extended quantum system's delocalization. This can be interpreted as if the particle is traversing a confined region with local temperature variations due to differences in acceleration. The spin thermometer essentially averages these local inverse temperatures, weighted by the particle's position within the system. It's essential to acknowledge the ``as if" nature of this interpretation: attributing a genuinely spatially varying temperature is inappropriate (as elucidated in Sec.  \ref{sec:ambiguity}); moreover, within the confines of conventional quantum mechanics, genuine particle motion finds no place. Nevertheless, this effectively encapsulates the non-local character of the phenomenon.

\begin{figure}
    \begin{center}
    \includegraphics[scale=.6]{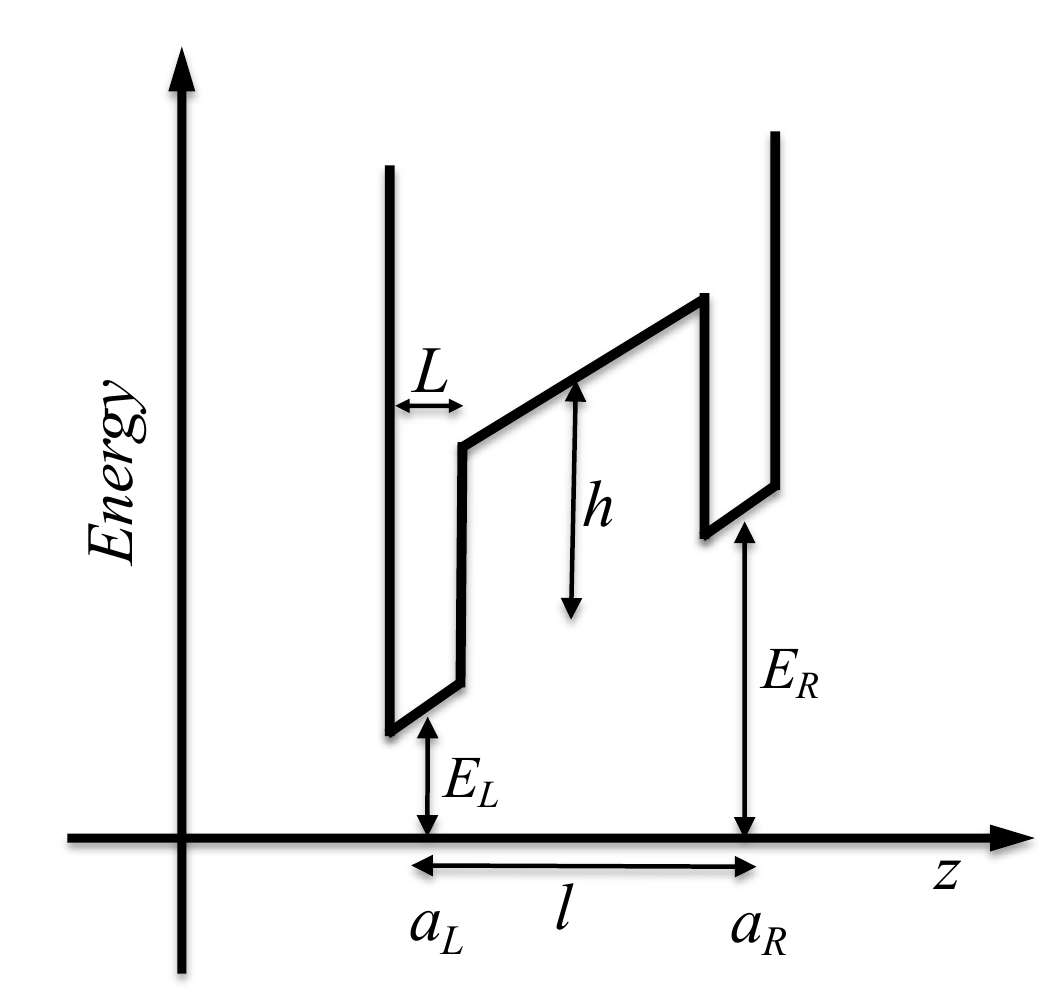}
       \end{center}
    \caption{
   A pictorial representation of the double well-confining potential.
    The left and right wells have length $L$ and are separated by a distance $l \gg L$.
    They are accelerated with proper acceleration $a_L$ and $a_R$, respectively.
    The symmetric potential is tilted because of the relativistic effects.
    } 
    \label{fig:potentials}
\end{figure} 

\subsection{Unruh effect in sharply localized multiple wells}
\label{sec:sharp_well}

Let us now consider a double-well confining potential.
For simplicity, we assume to have two square wells of linear dimension $L$ separated by $l$ as shown in Fig. \ref{fig:potentials}.
As in the previous section, we take \(a \sim 2.5 \times 10^{20}\) m/s\(^2\), and \(L \sim 10^{-7}\) m and assume that $L \ll l \lesssim 10^{-3}$~m~(extending to the upper limit the second condition in Eq. \eqref{eq:regime}), so that the system can be delocalized over a macroscopic distance while still remaining within the non-relativistic regime. This would further support the considerations made at the end of the previous section.

The large separation of the two wells implies that there is no tunneling between them.
For deep wells, only the lowest spatial eigenstates play a role in the dynamics, and these are strongly localized around the center of the wells.
At the same time, following the results obtained in Sec. \ref{sec:not_sharp_well}, for \(L \sim 10^{-7}\) m, we expect only the ground states associated with the left and right wells to be populated. We denote these eigenstates and eigenvalues with $\phi_{L,R}$ and $E_{L,R}$, respectively.
We also define the energy of the left (right) well in the inertial frame with $E_L^0$ ($E_R^0$) and the corresponding one with tilting due to the lapse with $E_L$ ($E_R$).

To have an estimate of the energy scales, we take the left well as the reference frame, i.e., $a=a_L$ and $\beta_a= \beta_L$, and rescale the energy to have $E_L=E_L^0=0$.
The right well ground state energy is $E_R = E_R^0 + m a l$ so that the relativistic tilting increases the energy of the well.

For a symmetric double well $E_R^0=E_L^0=0$, and $\beta_a E_R \sim 10^6$. 
Such value for the exponent implies that the right well is not populated, and, effectively, the particle is always found in the left well.
As a consequence, the spin state is thermal with inverse temperature $\beta_a=\beta_L$ even if the system has a macroscopic extension $l\sim 10^{-3}$~m.

However, there are situations, as discussed in Sec. \ref{sec:not_sharp_well}, where the spin-thermometer registers an average inverse temperature.
An example is the asymmetric well with $E_R^0 \sim -m a l$ so that $E_R  \sim E_L = 0$.
In this case, the relativistic tilting cancels the potential asymmetry, resulting in symmetric wells in the accelerated frame.
Again, only the ground state is populated, but it is extended over the two wells, i.e., $\Phi_g \sim (\Phi_L + \Phi_R)/\sqrt{2}$. Consequently,  from Eq. \eqref{eq:meanbeta}, the thermalizing inverse average temperature of the thermometer is 
\begin{equation}
    \overline{\beta}_0 = \frac{\beta_L +\beta_R}{2}
\end{equation}
with $\beta_R \sim \beta_L (1+a l/c^2)$.

\section{Conclusions}
\label{sec:conclusions}

We began by examining existing research on the Unruh effect in extended quantum systems. We then explored how the finite size of these systems becomes crucial. This analysis brings us closer to a realistic experimental setup, likely involving storage rings \cite{abi2021}, where the Unruh effect could potentially be observed---although for the latter case, further study of uniform circular acceleration is needed.

In this scenario, our thermometer is modeled as a single electron confined within a constantly accelerating box by external fields. This electron exists as a low-energy state of a zero-temperature fermionic field. The electron's spin interacts with a quantum field in its vacuum state, mimicking a thermal reservoir for the accelerated electron. While a true temperature cannot be definitively assigned to a delocalized electron, we can measure a unique value in a specific regime by carefully choosing the electron's spin as a temperature indicator.

This effect manifests itself most prominently when the electron's energy is low. Additionally, specific conditions are required to maintain its delocalization within the box. Under these circumstances, we can treat the electron's motion non-relativistically while retaining a remnant of relativistic effects through its spin degrees of freedom. This approach effectively captures the key aspects of the phenomenon without delving into all the complexities of a fully relativistic treatment.

Building on the arguments presented in Ref. \cite{Bell1985}, we show that the stationary solution for the reduced spin density matrix [see Eq. \eqref{eq:rether}] lacks a strictly thermal character. However, through a more refined approximation compared to Ref. \cite{Bell1985}, we identify and describe the precise conditions under which a form of thermalization can emerge, even for extended quantum systems. Our results indicate that under specific conditions, carefully designed measurements might lead to a reevaluation of our understanding of Unruh thermalization.  In particular, we analyzed two scenarios: a large confining well and sharply localized multiple wells. Notably, the calculated temperature in these cases represents an average across the entire confined region. This implies that while an extended system may not possess a single, well-defined temperature, specific measurements can still extract a unique effective temperature that reflects the nonlocal and extended nature of the quantum system.

\bigskip

\begin{acknowledgments}
The authors acknowledge financial support from INFN.
\end{acknowledgments}

\pagebreak
\widetext
\begin{center}
\textbf{\large Appendix}
\end{center}
\setcounter{section}{0}
\setcounter{equation}{0}
\setcounter{figure}{0}
\setcounter{table}{0}
\setcounter{page}{1}
\makeatletter
\renewcommand{\theequation}{S\arabic{equation}}
\renewcommand{\thefigure}{S\arabic{figure}}
\renewcommand{\bibnumfmt}[1]{[S#1]}
\renewcommand{\citenumfont}[1]{S#1}
\newtheorem{theo}{Theorem}

\section{Non-relativistic approximation of the Dirac equation in accelerated frames}
\label{app:non-relativistic_approximation of the Dirac equation}

The derivation of the non-relativistic approximation of Eq. \eqref{eq:hambas}  to yield  Eq. \eqref{eq:hambas} can be readily achieved by adapting standard methods (assuming $N(z)=1$) to the specific case at hand ($N(z) = 1 + {az}/{c^2}$). We include it here for completeness. To simplify the analysis, we will focus on the time-dependent  c-number Dirac equation
\begin{align}
\label{diraceq}
    i\hbar\dfrac{\de \Psi}{\de t} = \bigg[\bigg(1+\dfrac{az}{c^2}\bigg)(-ic\hbar\alpha^i\text{D}_i + mc^2\beta + e\phi) - i\dfrac{a\hbar}{2c}\alpha_z\bigg]\Psi.
\end{align}
where $\Psi = (\Bar{\chi},\bar{\phi})$ is a Dirac 4-spinor defined in terms of the 2-spinors  $\bar{\chi},\Bar{\phi}$. Accordingly, Eq. \eqref{diraceq} can be written as
\begin{align}
    i\hbar\dfrac{\de \bar{\chi}}{\de\taux } &= -i\hbar c\bigg(1+\frac{a\xix}{c^2}\bigg)\sigma\cdot \textbf{D} \bar{\phi} - i\hbar \dfrac{a}{2c}\sigma_z\bar{\phi} + \bigg(1+\frac{a\xix}{c^2}\bigg) (mc^2 + U(z))\bar{\chi} \\
    i\hbar\dfrac{\de \bar{\phi}}{\de\taux } &= -i\hbar c\bigg(1+\frac{a\xix}{c^2}\bigg)\sigma\cdot \textbf{D}\bar{\chi} - i\hbar \dfrac{a}{2c}\sigma_z\bar{\chi} -\bigg(1+\frac{a\xix}{c^2}\bigg)(mc^2 - U(z))\bar{\phi}.
\end{align}
Let $N(z)$  be the lapse function defined by Eq. \eqref{km}. Moreover, define   
\begin{equation}
\begin{pmatrix}
    \bar{\chi} \\ \bar{\phi}
\end{pmatrix} 
= \exp{\bigg[-i\frac{\taux  mc^2}{\hbar} N(z) \bigg]}\begin{pmatrix}
    \chi \\\phi
\end{pmatrix}.
\end{equation}
(see, e.g.,  \cite{Landau1991V3}). Then the 2-spinors $\phi$ and $\chi$ satisfy the equations
\begin{align}
\label{relschr}
    i\hbar\dfrac{\de \chi}{\de\taux } &= -i\hbar cN(z)\sigma\cdot \textbf{D} \phi - mac\taux  N(z)\sigma_z \phi - i\hbar \dfrac{a}{2c}\sigma_z\phi + N(z) U(\xix)\chi \\
    i\hbar\dfrac{\de \phi}{\de\taux } &= -i\hbar cN(z)\sigma\cdot \textbf{D}\chi - mac\taux  N(z) \sigma_z \chi - i\hbar \dfrac{a}{2c}\sigma_z\chi - N(z)(2mc^2 + U(z))\phi.
\end{align}
Assuming $ \hbar \de_\taux \phi \ll mc^2$ and $U(z) \ll mc^2$, the second equation becomes
\begin{equation}
    \phi \simeq -\dfrac{1}{2mc^2}\bigg(i\hbar c\sigma\cdot \textbf{D} + mac\taux \sigma_z + i\hbar \dfrac{a}{2c}\dfrac{1}{ N(z) }\sigma_z\bigg)\chi.
\end{equation}
Substituting the relation between $\phi$ and $\chi$ into Eq. \eqref{relschr}, it becomes
\begin{align*}
    i\hbar\dfrac{\de \chi}{\de\taux } &= N(z)\bigg(-\dfrac{\hbar^2}{2m}(\sigma\cdot \textbf{D})^2 +i \dfrac{\hbar a\taux }{2}(\sigma\cdot \textbf{D})\sigma_z - \dfrac{\hbar^2 a}{4mc^2}\dfrac{1}{N(z)}(\sigma\cdot \textbf{D})\sigma_z +\dfrac{\hbar^2 a^2}{4mc^4}\dfrac{1}{N(z)^2}\bigg)\chi \\
    &+ N(z) \sigma_z \bigg(i\dfrac{\hbar a\taux }{2} \sigma\cdot \textbf{D} + \dfrac{ma^2\taux ^2}{2}\sigma_z + i\dfrac{\hbar a^2 \taux }{4c^2} \dfrac{1}{N(z)}\sigma_z\bigg)\chi \\
    &+ \sigma_z\bigg(-\dfrac{\hbar^2 a}{4mc^2}\sigma\cdot \textbf{D} + i\dfrac{\hbar a^2\taux }{4c^2}\sigma_z - \dfrac{\hbar^2 a^2}{8mc^4}\dfrac{1}{N(z)}\sigma_z\bigg)\chi + N(z) U(\xix)\chi.
\end{align*}
Then, using $\{\sigma^i,\sigma^j\} = 2\delta^{ij}\mathbb{1}$, we obtain
\begin{align}
    \notag
    i\hbar\dfrac{\de \chi}{\de\taux } =~ &N(z)\bigg(-\dfrac{\hbar^2}{2m}(\sigma\cdot \textbf{D})^2 +i \hbar a\taux \delta^{iz}\text{D}_i\bigg)\chi
    + \dfrac{ma^2\taux ^2}{2}N(z) \chi \\
    \notag
    &+ \bigg(-\dfrac{\hbar^2 a}{2mc^2}\delta^{iz}\text{D}_i + i\dfrac{\hbar a^2\taux }{2c^2} + \dfrac{\hbar^2 a^2}{8mc^4}\dfrac{1}{N(z)}\bigg)\chi + N(z) U(\xix)\chi \\
    \label{chieq}
    i\hbar\dfrac{\de \chi}{\de\taux } =~ &\dfrac{1}{2m}N(z)\bigg(-i\hbar(\sigma\cdot \textbf{D}) - mat\sigma_z\bigg)^2\chi + \bigg(-\dfrac{\hbar^2 a}{2mc^2}\text{D}_z + i\dfrac{\hbar a^2\taux }{2c^2} + \dfrac{\hbar^2 a^2}{8mc^4}\dfrac{1}{N(z)}\bigg)\chi + N(z) U(z)\chi.
\end{align}
We now apply the transformation $\chi = e^{ima\xix \taux /\hbar}\Psi$ (with abuse of notation $\Psi $ is now denoting a 2-spinor). 
Thus, \eqref{chieq} becomes
\begin{align}
\label{app_eq:pauli_equation}
    i\hbar \dfrac{\de \Psi}{\de \taux } = -\dfrac{\hbar^2}{2m}N(z)(\sigma\cdot \textbf{D})^2\Psi -\dfrac{\hbar^2 a}{2mc^2}\text{D}_z\Psi + \dfrac{\hbar^2 a^2}{8mc^4}\dfrac{1}{N(z)}\Psi + N(z) U(z)\Psi + ma\xix\Psi.
\end{align}
Using the well know Pauli vector identity for two vectors $\textbf{u},\textbf{v}$
\begin{align*}
    (\sigma\cdot \textbf{u})(\sigma\cdot \textbf{v}) = \textbf{u}\cdot \textbf{v} + i \sigma\cdot (\textbf{u}\times \textbf{v})
\end{align*}
and given that
\begin{align*}
    (i\hbar \nabla + e\Avec)(i\hbar\nabla + e\Avec)\Psi = ie\hbar \textbf{B}\Psi
\end{align*}
where $\textbf{B} = \nabla\times\Avec$, the Pauli-like equation \eqref{app_eq:pauli_equation} becomes
\begin{align}
    i\hbar \dfrac{\de \Psi}{\de \taux } &= \dfrac{1}{2m}N(z)(i\hbar\nablavec + e \Avec)^2 \Psi - \dfrac{e\hbar}{2m}N(z)\sigmavec \cdot \Bvec \Psi \nonumber \\
    &+ \dfrac{i\hbar a}{2mc^2}\delta^{iz}(i\hbar\nablavec + e \Avec)_i \Psi + \dfrac{\hbar^2 a^2}{8mc^4}\dfrac{1}{N(z)} \Psi + N(z) U(z) \Psi + ma\xix \Psi.
    \label{eq_app:full_Pauli_equation}
\end{align}
At first order in $1/c$, Eq. \eqref{eq_app:full_Pauli_equation} is
\begin{align}
    i\hbar \dfrac{\de \Psi}{\de \taux } &= \dfrac{1}{2m}(i\hbar\nablavec + e \Avec)^2\Psi - \dfrac{e\hbar}{2m}\sigmavec \cdot \Bvec \Psi + U(z) \Psi + ma\xix \Psi
\end{align}
which is the same obtained in Ref. \cite{Bell1985}. 

However, in this approximation, the spin degrees of freedom, i.e., the ones used as a thermometer, are not affected by the lapse contribution that is of order $1/c^2$.
We can then consider the next expansion contribution, i.e., to the order $1/c^2$, and, to simplify the problem, we make the following additional assumptions:
$i)$  we assume that the magnetic field $B$ is strong enough to produce observable effects in the spin degrees of freedom and, $ii)$  that the confining potential $U(\xix)$ induces localization of the lowest eigenstates of the Hamiltonian in Eq. (\ref{eq_app:full_Pauli_equation}).
In this case, the equation becomes
\begin{align}
    i\hbar \dfrac{\de \Psi}{\de \taux } &= \dfrac{1}{2m}(i\hbar\nablavec + e \Avec)^2 \Psi + U(\xix) \Psi + ma\xix \Psi + N(z)\hbar \boldsymbol{\omega} \cdot \sigmavec \Psi
    \label{eq_app:Pauli_eq}
\end{align}
where we have substituted $\Bvec$ with the corresponding $\boldsymbol{\omega}= -e \textbf{B}/(2m)$. 
Note that, because we are considering a perturbative expansion on $1/c$, we are assuming $a\xix/c^2\ll 1$. Therefore, given that $a\approx 10^{20}\,\text{m}\cdot \text{sec}^{-2}$, the range of validity of Eq. \eqref{eq_app:Pauli_eq} is $\xix\ll 10^{-3}\,\text{m}$.

\section{Perturbative expansion of $H$}
\label{app:perturbative_expansion}

From Eq. (\ref{eq_app:Pauli_eq}), we obtain the Hamiltonian of the system $H$, respectively.
We denote with $\mathcal{E}_n^\sigma$ and $\ket{\Psi_n^\sigma}\ket{\sigma}$ (with $\sigma=\pm$) its eigenvalues and eigenstates.
With these notations, we can write the Hamiltonian $H$ as
\begin{equation}
    H = \sum_{n, \sigma}  \mathcal{E}_n^\sigma \ketbra{\Psi_n^\sigma\rangle |  \sigma}{\sigma| \langle \Psi_n^\sigma }.
    \label{eq:app_H_Sa}
\end{equation}
As discussed in the main text, we separate the Hamiltonian \eqref{eq:hambas} in three contributions: $H_{\text{space}}$, $H_{\text{spin}}$ and $H_{\text{int}}$ which include the spatial, the spin degrees of freedom and the interaction between the two.
Explicitly, they read
\begin{eqnarray}
    H_{\text{space}} &=& -\dfrac{\hbar^2}{2m} \frac{\partial^2}{\partial \xix^2} + U(\xix) + m a\xix \nonumber \\
    H_{\text{spin}} &=& \hbar \omega \sigma_z \nonumber \\
    H_{\text{int}} &=& \frac{a\xix}{c^2} \hbar \omega \sigma_z 
\end{eqnarray}
Projecting on the $\ket{\pm}$ spin sector, we obtain two eigenequations for the Hamiltonian \eqref{eq:hambas}
\begin{equation}
    H^\pm = -\dfrac{\hbar^2}{2m} \frac{\partial^2}{\partial \xix^2} + U(\xix) + m a\xix \pm \hbar \omega
\pm \frac{a\xix}{c^2} \hbar \omega
    \label{app_eq:H_space}
\end{equation}

Since in the realistic cases, $m c^2\gg \hbar \omega$, the last term associated to $H_{\text{int}}$ can be treated as a perturbation.
This allows us to separate the spatial and spin degrees of freedom and keep the first-order correction in $\hbar \omega/(m c^2)$ to obtain the eigenvalues and eigenstates \cite{Landau1991V3}
\begin{eqnarray}
    \mathcal{E}_n^\pm  &=& E_n \pm  (1+a \xix_n /c^2) \hbar \omega \nonumber \\
    |\Psi_n^{\pm_{(1)}}\rangle &= & \ket{\Phi_n \pm} \pm  \frac{\hbar a \omega}{c^2} \sum_{k}  \frac{\xix_n}{E_n-E_k} \ket{\Phi_k \pm},
    \label{app_eq:perturbed_eigs}
\end{eqnarray}
where $\xix_n = \matrixel{\Phi_n}{\xix}{\Phi_n}$, 
$E_n$ and $\ket{\Phi_n}$ are eigenvalues and eigenstates of the space Hamiltonian, respectively.
From this, we immediately obtain the stationary density matrix in Eq. \eqref{eq:spectral_rho_therm} in the main text.


\begin{thebibliography}{31}
\expandafter\ifx\csname natexlab\endcsname\relax\def\natexlab#1{#1}\fi
\expandafter\ifx\csname bibnamefont\endcsname\relax
  \def\bibnamefont#1{#1}\fi
\expandafter\ifx\csname bibfnamefont\endcsname\relax
  \def\bibfnamefont#1{#1}\fi
\expandafter\ifx\csname citenamefont\endcsname\relax
  \def\citenamefont#1{#1}\fi
\expandafter\ifx\csname url\endcsname\relax
  \def\url#1{\texttt{#1}}\fi
\expandafter\ifx\csname urlprefix\endcsname\relax\def\urlprefix{URL }\fi
\providecommand{\bibinfo}[2]{#2}
\providecommand{\eprint}[2][]{\url{#2}}

\bibitem[{\citenamefont{Bell et~al.}(1985)\citenamefont{Bell, Hughes, and
  Leinaas}}]{Bell1985}
\bibinfo{author}{\bibfnamefont{J.~S.} \bibnamefont{Bell}},
  \bibinfo{author}{\bibfnamefont{R.~J.} \bibnamefont{Hughes}},
  \bibnamefont{and} \bibinfo{author}{\bibfnamefont{J.~M.}
  \bibnamefont{Leinaas}}, \bibinfo{journal}{Zeitschrift f{\"u}r Physik C
  Particles and Fields} \textbf{\bibinfo{volume}{28}}, \bibinfo{pages}{75}
  (\bibinfo{year}{1985}).

\bibitem[{\citenamefont{Fulling}(1973)}]{Fulling1973}
\bibinfo{author}{\bibfnamefont{S.~A.} \bibnamefont{Fulling}},
  \bibinfo{journal}{Phys. Rev. D} \textbf{\bibinfo{volume}{7}},
  \bibinfo{pages}{2850} (\bibinfo{year}{1973}).

\bibitem[{\citenamefont{Davies}(1975)}]{Davies1975}
\bibinfo{author}{\bibfnamefont{P.~C.~W.} \bibnamefont{Davies}},
  \bibinfo{journal}{Journal of Physics A: Mathematical and General}
  \textbf{\bibinfo{volume}{8}}, \bibinfo{pages}{609} (\bibinfo{year}{1975}).

\bibitem[{\citenamefont{Unruh}(1976)}]{Unruh1976}
\bibinfo{author}{\bibfnamefont{W.~G.} \bibnamefont{Unruh}},
  \bibinfo{journal}{Phys. Rev. D} \textbf{\bibinfo{volume}{14}},
  \bibinfo{pages}{870} (\bibinfo{year}{1976}).

\bibitem[{\citenamefont{DeWitt}(1979)}]{DeWitt1979}
\bibinfo{author}{\bibfnamefont{B.~S.} \bibnamefont{DeWitt}},
  \emph{\bibinfo{title}{General relativity : an Einstein centenary survey /
  edited by S. W. Hawking, W. Israel}} (\bibinfo{publisher}{Cambridge
  University Press}, \bibinfo{address}{Cambridge (UK)}, \bibinfo{year}{1979}).

\bibitem[{\citenamefont{Bell and Leinaas}(1983)}]{Bell1983}
\bibinfo{author}{\bibfnamefont{J.}~\bibnamefont{Bell}} \bibnamefont{and}
  \bibinfo{author}{\bibfnamefont{J.}~\bibnamefont{Leinaas}},
  \bibinfo{journal}{Nuclear Physics B} \textbf{\bibinfo{volume}{212}},
  \bibinfo{pages}{131} (\bibinfo{year}{1983}).

\bibitem[{\citenamefont{Bell and Leinaas}(1987)}]{Bell1987}
\bibinfo{author}{\bibfnamefont{J.}~\bibnamefont{Bell}} \bibnamefont{and}
  \bibinfo{author}{\bibfnamefont{J.}~\bibnamefont{Leinaas}},
  \bibinfo{journal}{Nuclear Physics B} \textbf{\bibinfo{volume}{284}},
  \bibinfo{pages}{488} (\bibinfo{year}{1987}), ISSN \bibinfo{issn}{0550-3213}.

\bibitem[{\citenamefont{Leinaas}(2002)}]{Leinaas2002}
\bibinfo{author}{\bibfnamefont{J.~M.} \bibnamefont{Leinaas}},
  \emph{\bibinfo{title}{Quantum Aspects of Beam Physics}}
  (\bibinfo{publisher}{WORLD SCIENTIFIC}, \bibinfo{year}{2002}).

\bibitem[{\citenamefont{Abi et~al.}(2021)\citenamefont{Abi, Albahri, Al-Kilani,
  Allspach, Alonzi, Anastasi, Anisenkov, Azfar, Badgley, Bae\ss{}ler
  et~al.}}]{abi2021}
\bibinfo{author}{\bibfnamefont{B.}~\bibnamefont{Abi}},
  \bibinfo{author}{\bibfnamefont{T.}~\bibnamefont{Albahri}},
  \bibinfo{author}{\bibfnamefont{S.}~\bibnamefont{Al-Kilani}},
  \bibinfo{author}{\bibfnamefont{D.}~\bibnamefont{Allspach}},
  \bibinfo{author}{\bibfnamefont{L.~P.} \bibnamefont{Alonzi}},
  \bibinfo{author}{\bibfnamefont{A.}~\bibnamefont{Anastasi}},
  \bibinfo{author}{\bibfnamefont{A.}~\bibnamefont{Anisenkov}},
  \bibinfo{author}{\bibfnamefont{F.}~\bibnamefont{Azfar}},
  \bibinfo{author}{\bibfnamefont{K.}~\bibnamefont{Badgley}},
  \bibinfo{author}{\bibfnamefont{S.}~\bibnamefont{Bae\ss{}ler}},
  \bibnamefont{et~al.} (\bibinfo{collaboration}{Muon $g\ensuremath{-}2$
  Collaboration}), \bibinfo{journal}{Phys. Rev. Lett.}
  \textbf{\bibinfo{volume}{126}}, \bibinfo{pages}{141801}
  (\bibinfo{year}{2021}).

\bibitem[{\citenamefont{Mart\'{\i}n-Mart\'{\i}nez
  et~al.}(2011)\citenamefont{Mart\'{\i}n-Mart\'{\i}nez, Fuentes, and
  Mann}}]{Martin-Martinez2011}
\bibinfo{author}{\bibfnamefont{E.}~\bibnamefont{Mart\'{\i}n-Mart\'{\i}nez}},
  \bibinfo{author}{\bibfnamefont{I.}~\bibnamefont{Fuentes}}, \bibnamefont{and}
  \bibinfo{author}{\bibfnamefont{R.~B.} \bibnamefont{Mann}},
  \bibinfo{journal}{Phys. Rev. Lett.} \textbf{\bibinfo{volume}{107}},
  \bibinfo{pages}{131301} (\bibinfo{year}{2011}).

\bibitem[{\citenamefont{\ifmmode~\check{S}\else \v{S}\fi{}oda
  et~al.}(2022)\citenamefont{\ifmmode~\check{S}\else \v{S}\fi{}oda, Sudhir, and
  Kempf}}]{Soda2022}
\bibinfo{author}{\bibfnamefont{B.}~\bibnamefont{\ifmmode~\check{S}\else
  \v{S}\fi{}oda}}, \bibinfo{author}{\bibfnamefont{V.}~\bibnamefont{Sudhir}},
  \bibnamefont{and} \bibinfo{author}{\bibfnamefont{A.}~\bibnamefont{Kempf}},
  \bibinfo{journal}{Phys. Rev. Lett.} \textbf{\bibinfo{volume}{128}},
  \bibinfo{pages}{163603} (\bibinfo{year}{2022}).

\bibitem[{\citenamefont{Stargen and Lochan}(2022)}]{Stargen2022}
\bibinfo{author}{\bibfnamefont{D.~J.} \bibnamefont{Stargen}} \bibnamefont{and}
  \bibinfo{author}{\bibfnamefont{K.}~\bibnamefont{Lochan}},
  \bibinfo{journal}{Phys. Rev. Lett.} \textbf{\bibinfo{volume}{129}},
  \bibinfo{pages}{111303} (\bibinfo{year}{2022}).

\bibitem[{\citenamefont{Gooding et~al.}(2020)\citenamefont{Gooding, Biermann,
  Erne, Louko, Unruh, Schmiedmayer, and Weinfurtner}}]{Gooding2020}
\bibinfo{author}{\bibfnamefont{C.}~\bibnamefont{Gooding}},
  \bibinfo{author}{\bibfnamefont{S.}~\bibnamefont{Biermann}},
  \bibinfo{author}{\bibfnamefont{S.}~\bibnamefont{Erne}},
  \bibinfo{author}{\bibfnamefont{J.}~\bibnamefont{Louko}},
  \bibinfo{author}{\bibfnamefont{W.~G.} \bibnamefont{Unruh}},
  \bibinfo{author}{\bibfnamefont{J.}~\bibnamefont{Schmiedmayer}},
  \bibnamefont{and}
  \bibinfo{author}{\bibfnamefont{S.}~\bibnamefont{Weinfurtner}},
  \bibinfo{journal}{Phys. Rev. Lett.} \textbf{\bibinfo{volume}{125}},
  \bibinfo{pages}{213603} (\bibinfo{year}{2020}).

\bibitem[{\citenamefont{Buchholz and Verch}(2015)}]{Buchholz_2015}
\bibinfo{author}{\bibfnamefont{D.}~\bibnamefont{Buchholz}} \bibnamefont{and}
  \bibinfo{author}{\bibfnamefont{R.}~\bibnamefont{Verch}},
  \bibinfo{journal}{Classical and Quantum Gravity}
  \textbf{\bibinfo{volume}{32}}, \bibinfo{pages}{245004}
  (\bibinfo{year}{2015}).

\bibitem[{\citenamefont{Korsbakken and Leinaas}(2004)}]{Korsbakken_2004}
\bibinfo{author}{\bibfnamefont{J.~I.} \bibnamefont{Korsbakken}}
  \bibnamefont{and} \bibinfo{author}{\bibfnamefont{J.~M.}
  \bibnamefont{Leinaas}}, \bibinfo{journal}{Phys. Rev. D}
  \textbf{\bibinfo{volume}{70}}, \bibinfo{pages}{084016}
  (\bibinfo{year}{2004}).

\bibitem[{\citenamefont{Lima et~al.}(2019)\citenamefont{Lima, Brito, Hoyos, and
  Vanzella}}]{Lima2019}
\bibinfo{author}{\bibfnamefont{C.~A.~U.} \bibnamefont{Lima}},
  \bibinfo{author}{\bibfnamefont{F.}~\bibnamefont{Brito}},
  \bibinfo{author}{\bibfnamefont{J.}~\bibnamefont{Hoyos}}, \bibnamefont{and}
  \bibinfo{author}{\bibfnamefont{D.~A.~T.} \bibnamefont{Vanzella}},
  \bibinfo{journal}{Nature Communications} \textbf{\bibinfo{volume}{10}},
  \bibinfo{pages}{3030} (\bibinfo{year}{2019}).

\bibitem[{\citenamefont{Foo et~al.}(2020)\citenamefont{Foo, Onoe, and
  Zych}}]{Foo2020}
\bibinfo{author}{\bibfnamefont{J.}~\bibnamefont{Foo}},
  \bibinfo{author}{\bibfnamefont{S.}~\bibnamefont{Onoe}}, \bibnamefont{and}
  \bibinfo{author}{\bibfnamefont{M.}~\bibnamefont{Zych}},
  \bibinfo{journal}{Phys. Rev. D} \textbf{\bibinfo{volume}{102}},
  \bibinfo{pages}{085013} (\bibinfo{year}{2020}).

\bibitem[{\citenamefont{Barbado et~al.}(2020)\citenamefont{Barbado,
  Castro-Ruiz, Apadula, and Brukner}}]{Barbado2020}
\bibinfo{author}{\bibfnamefont{L.~C.} \bibnamefont{Barbado}},
  \bibinfo{author}{\bibfnamefont{E.}~\bibnamefont{Castro-Ruiz}},
  \bibinfo{author}{\bibfnamefont{L.}~\bibnamefont{Apadula}}, \bibnamefont{and}
  \bibinfo{author}{\bibfnamefont{i.~c.~v.} \bibnamefont{Brukner}},
  \bibinfo{journal}{Phys. Rev. D} \textbf{\bibinfo{volume}{102}},
  \bibinfo{pages}{045002} (\bibinfo{year}{2020}).

\bibitem[{\citenamefont{M\o{}ller}(1972)}]{Moller1972}
\bibinfo{author}{\bibfnamefont{C.}~\bibnamefont{M\o{}ller}},
  \emph{\bibinfo{title}{The Theory of Relativity}}
  (\bibinfo{publisher}{Clarendon Press}, \bibinfo{address}{Oxford, UK},
  \bibinfo{year}{1972}).

\bibitem[{\citenamefont{Kottler}(1914)}]{Kottler1914}
\bibinfo{author}{\bibfnamefont{F.}~\bibnamefont{Kottler}},
  \bibinfo{journal}{Annalen der Physik} \textbf{\bibinfo{volume}{349}},
  \bibinfo{pages}{701} (\bibinfo{year}{1914}).

\bibitem[{\citenamefont{Kubo}(1957)}]{Kubo1957}
\bibinfo{author}{\bibfnamefont{R.}~\bibnamefont{Kubo}},
  \bibinfo{journal}{Journal of the Physical Society of Japan}
  \textbf{\bibinfo{volume}{12}}, \bibinfo{pages}{570} (\bibinfo{year}{1957}).

\bibitem[{\citenamefont{Martin and Schwinger}(1959)}]{Martin-Schwinger1959}
\bibinfo{author}{\bibfnamefont{P.~C.} \bibnamefont{Martin}} \bibnamefont{and}
  \bibinfo{author}{\bibfnamefont{J.}~\bibnamefont{Schwinger}},
  \bibinfo{journal}{Phys. Rev.} \textbf{\bibinfo{volume}{115}},
  \bibinfo{pages}{1342} (\bibinfo{year}{1959}).

\bibitem[{\citenamefont{Sewell}(1982)}]{SEWELL1982}
\bibinfo{author}{\bibfnamefont{G.~L.} \bibnamefont{Sewell}},
  \bibinfo{journal}{Annals of Physics} \textbf{\bibinfo{volume}{141}},
  \bibinfo{pages}{201} (\bibinfo{year}{1982}).

\bibitem[{\citenamefont{Breuer and Petruccione}(2002)}]{Breuer_book}
\bibinfo{author}{\bibfnamefont{H.~P.} \bibnamefont{Breuer}} \bibnamefont{and}
  \bibinfo{author}{\bibfnamefont{F.}~\bibnamefont{Petruccione}},
  \emph{\bibinfo{title}{The theory of open quantum systems}}
  (\bibinfo{publisher}{Oxford University Press}, \bibinfo{address}{Oxford, UK},
  \bibinfo{year}{2002}).

\bibitem[{\citenamefont{Blum}(2012)}]{Blum_book}
\bibinfo{author}{\bibfnamefont{K.}~\bibnamefont{Blum}},
  \emph{\bibinfo{title}{Density matrix theory and applications}}
  (\bibinfo{publisher}{Springer}, \bibinfo{address}{Berlin},
  \bibinfo{year}{2012}).

\bibitem[{\citenamefont{Goldstein et~al.}(2010)\citenamefont{Goldstein,
  Lebowitz, Mastrodonato, Tumulka, and Zanghi}}]{Goldstein2010}
\bibinfo{author}{\bibfnamefont{S.}~\bibnamefont{Goldstein}},
  \bibinfo{author}{\bibfnamefont{J.~L.} \bibnamefont{Lebowitz}},
  \bibinfo{author}{\bibfnamefont{C.}~\bibnamefont{Mastrodonato}},
  \bibinfo{author}{\bibfnamefont{R.}~\bibnamefont{Tumulka}}, \bibnamefont{and}
  \bibinfo{author}{\bibfnamefont{N.}~\bibnamefont{Zanghi}},
  \bibinfo{journal}{Phys. Rev. E} \textbf{\bibinfo{volume}{81}},
  \bibinfo{pages}{011109} (\bibinfo{year}{2010}).

\bibitem[{\citenamefont{Srednicki}(1994)}]{Srednicki1994}
\bibinfo{author}{\bibfnamefont{M.}~\bibnamefont{Srednicki}},
  \bibinfo{journal}{Phys. Rev. E} \textbf{\bibinfo{volume}{50}},
  \bibinfo{pages}{888} (\bibinfo{year}{1994}).

\bibitem[{\citenamefont{Thaller}(1992)}]{thaller}
\bibinfo{author}{\bibfnamefont{B.}~\bibnamefont{Thaller}},
  \emph{\bibinfo{title}{{The Dirac equation}}} (\bibinfo{address}{Berlin},
  \bibinfo{year}{1992}).

\bibitem[{\citenamefont{Bjorken and Drell}(1965)}]{bjorken}
\bibinfo{author}{\bibfnamefont{J.}~\bibnamefont{Bjorken}} \bibnamefont{and}
  \bibinfo{author}{\bibfnamefont{S.}~\bibnamefont{Drell}},
  \emph{\bibinfo{title}{Relativistic Quantum Fields}}
  (\bibinfo{publisher}{McGraw-Hill}, \bibinfo{address}{New York, NY},
  \bibinfo{year}{1965}).

\bibitem[{\citenamefont{Weinberg}(2005)}]{weinberg}
\bibinfo{author}{\bibfnamefont{S.}~\bibnamefont{Weinberg}},
  \emph{\bibinfo{title}{The Quantum Theory of Fields, Volume 1: Foundations}}
  (\bibinfo{publisher}{Cambridge University Press, Cambridge, UK},
  \bibinfo{year}{2005}), ISBN \bibinfo{isbn}{0521670535}.

\bibitem[{\citenamefont{Landau and Lifshits}(1991)}]{Landau1991V3}
\bibinfo{author}{\bibfnamefont{L.~D.} \bibnamefont{Landau}} \bibnamefont{and}
  \bibinfo{author}{\bibfnamefont{E.~M.} \bibnamefont{Lifshits}},
  \emph{\bibinfo{title}{Quantum Mechanics: Non-Relativistic Theory}}, vol.
  \bibinfo{volume}{v.3} of \emph{\bibinfo{series}{Course of Theoretical
  Physics}} (\bibinfo{publisher}{Butterworth-Heinemann},
  \bibinfo{address}{Oxford}, \bibinfo{year}{1991}), ISBN
  \bibinfo{isbn}{978-0-7506-3539-4}.

\end{thebibliography}
\end{document}